\documentclass[a4paper]{cas-sc}
\usepackage[square,numbers]{natbib}
\usepackage{graphics}
\usepackage{microtype}
\usepackage{siunitx}
\usepackage{mathtools}
\usepackage{amsmath} 
\usepackage{lipsum}

\newcolumntype{Y}{>{\centering\arraybackslash}X}

\begin{document}

\title[mode=title]{Unsupervised Machine Learning of the Contact Process}

\shorttitle{Unsupervised Machine Learning of the Contact Process}

\shortauthors{L. T. Brito et~al.}

\author[1,2]{L. T. Brito}
\credit{Research Development, Methodology, Programming, Data Curation}

\author[2]{G. A. Alves}
\credit{Research Development, Methodology, Writing - review \& editing}

\author[1]{F. W. S. Lima}
\credit{Research Development, Methodology, Writing - review \& editing}

\author[2]{A. Macedo-Filho}
\credit{Research Development, Methodology, Writing - review \& editing}

\author[3]{R. S. Ferreira}
\credit{Research Development, Methodology, Writing - review \& editing}

\author[1]{T. F. A. Alves}
[orcid=0000-0001-9937-7980]
\cormark[1]
\ead{tay@ufpi.edu.br}
\credit{Research Development, Methodology, Data curation, Writing - Original draft}

\affiliation[2]{organization={Universidade Estadual do Piau\'{i}},
   addressline={Departamento de F\'{i}sica}, 
   city={Teresina},
   postcode={64002-150}, 
   state={Piau\'{i}},
   country={Brazil}
}

\affiliation[1]{organization={Universidade Federal do Piau\'{i}},
   addressline={Departamento de F\'{i}sica}, 
   city={Teresina},
   postcode={64049-550}, 
   state={Piau\'{i}},
   country={Brazil}
}

\affiliation[3]{organization={Universidade Federal de Ouro Preto},
   addressline={Departamento de Ci\^{e}ncias Exatas e Aplicadas}, 
   city={Jo\~{a}o Monlevade},
   postcode={35931-008}, 
   state={Minas Gerais},
   country={Brazil}
}

\date{Received: date / Revised version: date}

\begin{abstract}
   We investigate how unsupervised machine-learning methods can characterize the active-absorbing phase transition in the contact process in one and two spatial dimensions. Our analysis focuses on principal component analysis (PCA) and variational autoencoders (VAEs), and we show that direct applications of these methods encounter a central difficulty: contact process configurations are intrinsically non-negative. In the PCA case, standard centering procedures fail to isolate the order parameter in the leading component. To address this, we augment the data with sign-reversed configurations, creating a balanced dataset with zero mean while preserving the density information required to describe the transition. This preprocessing allows the dominant principal component to recover the order parameter and reproduce the expected finite-size scaling near criticality. For VAEs, we find a related limitation: a fixed Gaussian prior does not provide sufficient latent regularization across the full range of control parameters and system sizes for non-negative data. We overcome this by adopting a heteroscedastic Gaussian prior that adapts to the control parameter, leading to a substantial improvement in the finite-size scaling of reconstruction-based observables near the critical threshold. Taken together, these results highlight a key lesson: the success of machine learning in phase-transition problems does not rely on algorithmic complexity alone, but on tailoring the representation to respect the physical structure of the system, including its intrinsic constraints and symmetries.
\end{abstract}


\ExplSyntaxOn
\keys_set:nn { stm / mktitle } { nologo }
\ExplSyntaxOff

\maketitle

\section{Introduction}

Far from equilibrium phase transitions are ubiquitous in nature, manifesting through collective behavior, critical fluctuations, and universal scaling laws even without an underlying thermodynamic description. A particularly striking example is the active-absorbing phase transition found in systems with absorbing states, which are states from which the dynamics cannot naturally escape. The contact process exemplifies this behavior: it combines simple microscopic rules with direct relevance to real epidemic spreading, making it a canonical model for studying such transitions. Importantly, the continuous phase transition in the contact process belongs to the directed percolation universality class, whose scaling properties have been thoroughly mapped across different spatial dimensions~\cite{jensen-1992,hinrichsen-2000,henkel-2008}.

In the contact process, each lattice site is represented by a binary occupation variable, with infected (active) sites spontaneously becoming inactive at a unit rate and inactive sites being activated through contact with one of the active neighbors randomly chosen at a rate controlled by the basic reproduction number $\lambda$. The system undergoes a continuous transition at $\lambda=\lambda_c$ between an absorbing phase, in which the stationary density of active sites vanishes, and an active phase with a finite density. The stationary density $\rho$ therefore plays the role of the order parameter. Close to criticality, the observables obey finite-size and critical scaling laws governed by the directed percolation exponents~\cite{hinrichsen-2000,odor-2004,henkel-2008}. Although the underlying stochastic rules are local and simple, there is no exact solution even in the one-dimensional chain.

Machine-learning methods offer an alternative, data-driven route to the characterization of the phase transition. In statistical physics, supervised neural networks have been shown to classify phases and locate critical thresholds from raw configurations, while unsupervised methods can reveal low-dimensional structures without explicit phase labels or prior knowledge of the order parameter~\cite{carrasquilla-2017,mehta-2019}. Principal component analysis (PCA), in particular, has been used to recover quantities closely related to the order parameter in equilibrium spin models~\cite{wang-2016,wetzel-2017,hu-2017}. 

Moreover, variational autoencoders (VAEs) provide a probabilistic approach, in which an encoder maps microscopic configurations to a latent distribution and a decoder reconstructs, or generates, configurations from the latent space. In several lattice models, the latent representation or the reconstruction loss develops characteristic signatures across a phase transition~\cite{wetzel-2017,yevick-2022}. Variational autoencoders have been successfully employed to characterize nonequilibrium phase transitions~\cite{neto-2025, neto-2026}, where the VAE identified the phase transition through both the reconstruction loss and the correlation between real and reconstructed states. In particular, the latter is consistent with a universal quantity at criticality and exhibits finite-size scaling governed by the correlation-length exponent. These results indicate that VAEs can successfully capture the critical structure of nonequilibrium systems whose configuration space possesses an inversion symmetry.

Extending these ideas to absorbing-state transitions is appealing, but it introduces additional difficulties. Contact process configurations are non-negative and therefore lack the natural centering mechanism present in the Ising model, where spin inversion can be used to construct a centered dataset by augmenting it with sign-reversed configurations. Nevertheless, machine-learning studies of directed percolation and related processes with absorbing states have also demonstrated that both supervised and unsupervised architectures can locate transition thresholds and extract relevant features from spatiotemporal or steady-state configurations~\cite{shen-2021,shen-2022,gao-2025}. These results also indicate, however, that the physical interpretation of a learned coordinate depends sensitively on how the microscopic states are represented. A similar dependence on representation has been observed in unsupervised studies of multistate spin models, where an encoding adapted to the symmetry of the configuration space can substantially improve the PCA projection~\cite{tirelli-2022}. These observations motivate the preprocessing step adopted here, which preserves the density direction while producing an exactly centered ensemble.

Here we introduce such a construction by augmenting every physical configuration $\boldsymbol{\eta}$ with artificial sign-reversed configurations $-\boldsymbol{\eta}$. The transformation is used only in data space and does \emph{not} represent a physical symmetry of the contact process. Because the augmented ensemble contains the pair $\{\boldsymbol{\eta},-\boldsymbol{\eta}\}$ for each sample, its mean vanishes identically, while the covariance retains the information carried by the magnitude and spatial structure of the original configurations. We show that this simple doubling of the data qualitatively changes the PCA representation: the leading component becomes directly related to the non-negative order parameter. Moreover, PCA-based quantities display the finite-size scaling expected close to the critical threshold. This provides a concrete example in which a physically informed representation, rather than a more complex learning algorithm, is the essential ingredient for recovering an interpretable collective variable.

A related issue arises in generative models: the contact process is qualitatively different from systems with a built-in $\mathbb{Z}_2$ symmetry, because the configuration space is intrinsically non-symmetric. This asymmetry motivates replacing the standard fixed Gaussian prior of a VAE by an adaptive prior. In the usual formulation, the VAE relies on a fixed Gaussian prior, typically a parameter-independent standard normal distribution. However, for the architecture and training procedure used here, a standard Gaussian prior does not provide adequate latent regularization across the full range of $\lambda$ and system sizes. A fixed prior is therefore too restrictive to capture the relevant structure of the dataset. Indeed, generative studies of lattice systems have shown that visually plausible reconstructions do not, by themselves, guarantee that the generated ensemble preserves the physically relevant correlations and constraints~\cite{yevick-2022}.

To address the need for an adaptive prior, we replace the fixed VAE prior by a $\lambda$-dependent heteroscedastic Gaussian prior,
\begin{equation}
   p_{\phi}(z\mid\lambda) = \mathcal{N}\!\left(\mu_{p}(\lambda),\sigma_{p}^{2}(\lambda)\right),
\end{equation}
whose mean and variance are learned as functions of the control parameter. With this conditional prior, the latent distribution is allowed to adapt continuously to the active-site configuration dataset. We find that the normalized correlation between the input configurations and the configurations reconstructed by the heteroscedastic VAE is consistent with an universal value at the critical threshold, whose crossing point for different lattice sizes can be used to locate $\lambda_c$. The corresponding behavior is not recovered with a fixed Gaussian prior. Thus, in both the linear PCA analysis and the nonlinear generative model, the decisive step is to modify the statistical representation so that it respects the asymmetric, non-negative structure of the contact process data.

In this work, we investigate these effects for the contact process in the two-dimensional square lattice. Our main results are twofold. First, we demonstrate that sign-reversed data augmentation enables PCA to reconstruct the order parameter and reproduce the expected critical scaling. Second, we show that a heteroscedastic VAE can identify the transition. Taken together, these results emphasize that unsupervised machine learning of nonequilibrium criticality is not determined solely by the expressive power of an algorithm: the representation of the data and the inductive assumptions built into the latent distribution can be equally important.

The paper is organized as follows. We first introduce the contact process dynamics, simulation protocol, and finite-size observables. We then discuss the PCA construction and the effect of the sign-reversed data augmentation in Sec.~\ref{sec:pca_results}. Next, we present the variational autoencoder architecture and the heteroscedastic prior, followed by the correlation and scaling analyses in Sec.~\ref{sec:vae_results}. Finally, we summarize the implications of these results for machine-learning approaches to absorbing-state phase transitions in Sec.~\ref{sec:conclusions}.

\section{Data Generation}\label{sec:data_generation}

The contact process is a paradigmatic stochastic lattice model that exhibits a continuous nonequilibrium phase transition from an active phase to an absorbing state~\cite{harris-1974,hinrichsen-2000,odor-2004,henkel-2008}. Each lattice site is described by a binary variable $\eta_i$, where $\eta_i=0$ and $\eta_i=1$ denote susceptible and infected sites, respectively. The dynamics is governed by the control parameter $\lambda$, which sets the infection rate relative to the recovery rate.

We employ a random-sequential update scheme. During each elementary update, a site $i$ is selected uniformly at random.
\begin{itemize}
   \item If the selected site is infected, it becomes susceptible with recovery probability $p_{\mathrm{rec}} = \min\left(\lambda^{-1},1\right)$;
   \item If the selected site is susceptible, one of its nearest neighbors, denoted by $j$, is chosen uniformly at random. If this neighbor is infected, site $i$ becomes infected with probability $p_{\mathrm{inf}} = \min(\lambda,1)$. If the selected neighbor is susceptible, the configuration remains unchanged.
\end{itemize}
This prescription ensures that both transition probabilities, $p_{\mathrm{rec}}$ and $p_{\mathrm{inf}}$, remain bounded by unity while preserving their ratio, $p_{\mathrm{inf}}/p_{\mathrm{rec}}=\lambda$. One Monte Carlo step (MCS), which defines one unit of simulation time, consists of $N$ elementary update attempts.

Depending on $\lambda$, the model exhibits two stationary regimes. For $\lambda<\lambda_c$, the system is in the absorbing phase, in which the activity eventually vanishes. This phase is characterized by absorbing configurations: once the dynamics reaches one of these configurations, it cannot leave it. For $\lambda>\lambda_c$, in the idealized case of an infinite lattice, activity survives indefinitely in the thermodynamic limit, and the system is in the active phase. The continuous transition separating these two regimes belongs to the directed percolation universality class~\cite{hinrichsen-2000,odor-2004,henkel-2008}.

In finite systems, however, fluctuations can drive the dynamics into the absorbing configuration even in the active regime. To prevent the simulation from becoming permanently trapped in this state, we employ a reactivation (or resurrection) dynamics~\cite{pruessner-2007,macedo-filho-2018}. Whenever the system reaches the absorbing configuration, a single site is activated at random, and the stochastic evolution is resumed. This procedure enables the continuous sampling of stationary active configurations throughout the simulations.

The simulations were performed on lattices with periodic boundary conditions. In two dimensions, we considered the square lattice. A system configuration is represented by
\begin{equation}
   \boldsymbol{\eta} = (\eta_1,\eta_2,\ldots,\eta_N),
\label{eq:configuration}
\end{equation}
where $N$ denotes the number of sites. For a two-dimensional lattice with linear size $L$, we have $N=L^2$. The configurations $\boldsymbol{\eta}$ generated by the stochastic dynamics constitute the input data for the machine-learning analyses.

For each value of the control parameter, the system was first evolved for $10^6$ Monte Carlo steps before data collection began. Subsequent configurations were sampled at intervals of $10^3$ Monte Carlo steps to reduce correlations between consecutive samples. We considered the linear system sizes $L=16$, $20$, $24$, $32$, $40$, $48$, $56$, and $64$.

The infection rate $\lambda$ was sampled over the interval $[0.5\lambda_c,1.5\lambda_c]$, yielding a balanced dataset with configurations from both the absorbing and active phases. For a sufficiently broad interval, the method is relatively insensitive to the initial estimate of $\lambda_c$ used to define this range. This estimate can be progressively refined through successive iterations of the training procedure, each based on a more accurately balanced dataset centered on the updated critical-point estimate. Thus, an initial estimate obtained from standard methods, such as cumulant analysis, is sufficient to initialize the procedure.

For the training process, $200$ values of $\lambda$ were used, while an additional $100$ values from the same interval were reserved for inference. For each system size, the complete training set contained $N_{\mathcal D}=2\times10^6$ configurations. The dataset was balanced, with $N_{\mathcal D}/2$ configurations associated with each phase. By contrast, the PCA and variational autoencoder analyses discussed below were performed without phase labels.

\section{Principal component analysis}\label{sec:pca_results}

As discussed above, a straightforward application of PCA to the original non-negative configurations does not reproduce the order parameter.However, augmenting the dataset by including for every configuration $\boldsymbol{\eta}$, its sign-reversed counterpart $-\boldsymbol{\eta}$ allows the principal component to recover the order parameter. Thus, for a given system size, the dataset used in the PCA can be written schematically as
\begin{equation}
   {\cal D}_{\rm PCA} = \left\{ \boldsymbol{\eta}^{(1)},\ldots,\boldsymbol{\eta}^{(N_{\cal D})},
                               -\boldsymbol{\eta}^{(1)},\ldots,-\boldsymbol{\eta}^{(N_{\cal D})}
                        \right\}.
   \label{eq:pca_augmented_dataset}
\end{equation}
This construction yields a vanishing sample mean while preserving the information contained in the original configurations. Notice that the data augmentation introduced in Eq.~\eqref{eq:pca_augmented_dataset} is only a preprocessing operation in configuration space and does not represent a physical symmetry of the contact process. The principal component analysis was performed in Python using the \texttt{sklearn.decomposition.PCA} implementation of the \texttt{scikit-learn} library~\cite{pedregosa-2011}.

Let $\Lambda_k$ and $\boldsymbol{v}_k$ denote, respectively, the eigenvalues and eigenvectors of the covariance matrix in the decreasing order. The projection of a configuration onto the $k$th principal direction is
\begin{equation}
   p_k = \boldsymbol{v}_k^{\,T}\boldsymbol{\eta} = \sum_{i=1}^{N} v_{k,i}\eta_i,
\end{equation}
where $\boldsymbol{v}_k^{\,T}$ denotes the transpose of the $k$th principal eigenvector $\boldsymbol{v}_k$ of the covariance matrix.

\begin{figure}[pos=h!]
   \begin{center}
      \includegraphics[scale=0.4]{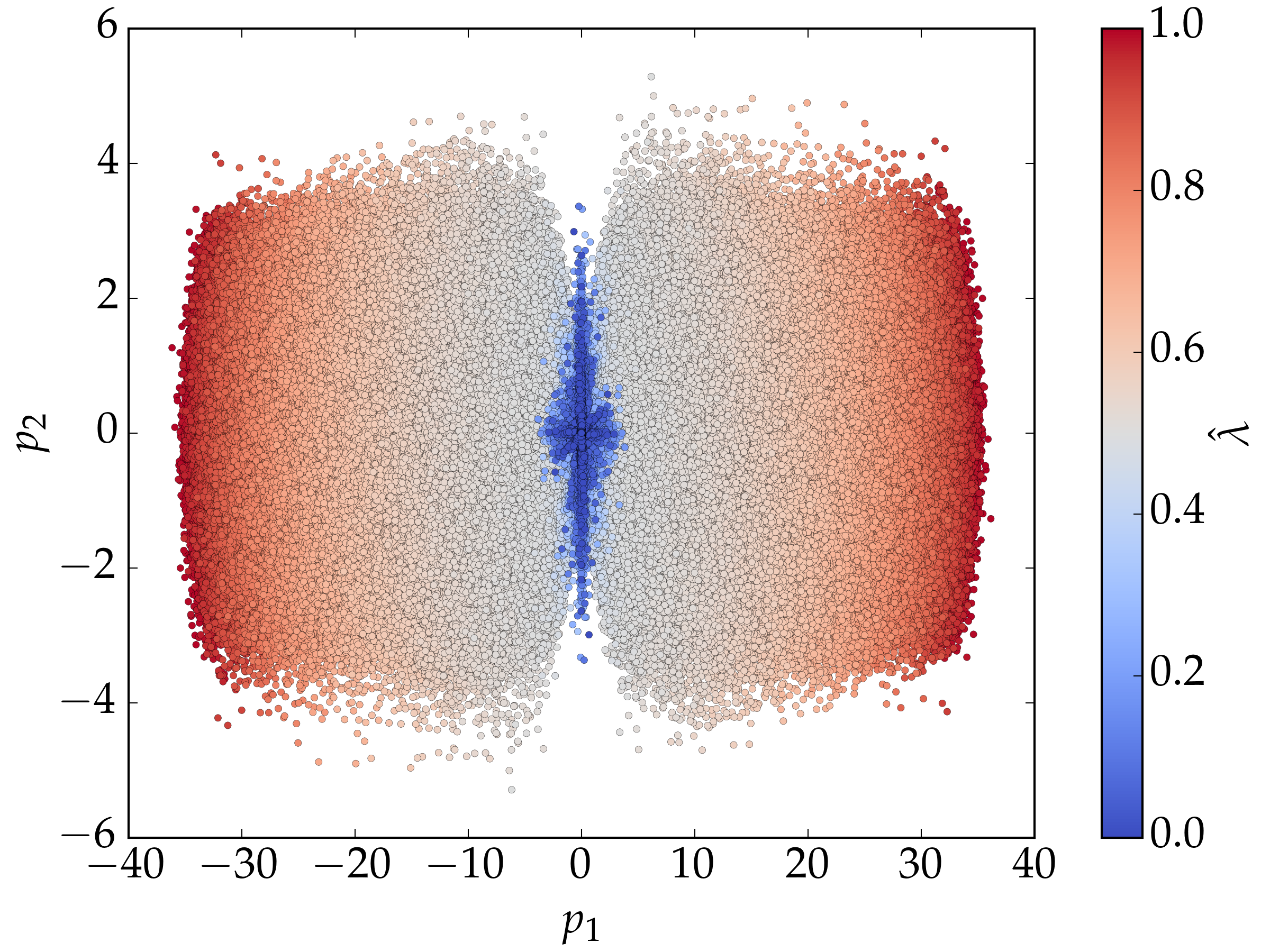}
   \end{center}
   \caption{Scatter plot of the two leading principal components, $p_1$ and $p_2$, for the contact process on the square lattice with $L=64$. The color scale represents the rescaled control parameter $\hat{\lambda}\in[0,1]$, corresponding to $\lambda\in[0.5\lambda_c,1.5\lambda_c]$. The compact cloud around the origin corresponds to configurations close to the absorbing state, whereas the two extended sign-related branches are associated predominantly with the active phase and the two-branch structure are due to the sign-reversed data augmentation used in the PCA analysis.}
   \label{fig:pca_scatter}
\end{figure}

Fig.~\ref{fig:pca_scatter} shows the projections onto the two leading principal components, $p_1$ and $p_2$, for $L=64$. Each point represents a microscopic configuration, colored according to the rescaled parameter $\hat{\lambda}$, which maps $\lambda\in[0.5\lambda_c,1.5\lambda_c]$ onto $[0,1]$. At small $\hat{\lambda}$, configurations form a narrow cluster near the origin, reflecting the low activity of the absorbing regime. As $\lambda$ approaches and exceeds the critical region, the distribution broadens and develops two approximately sign-related branches along $p_1$. These outer clouds, mainly associated with the active phase, possess a two-branch structure resulting from the sign-reversed data augmentation. Although this two-branch structure is artificial, the transition from a compact central cluster to an extended distribution clearly signals the active-absorbing phase transition.

Recent work \cite{muzzi-2024} has shown that uncentered PCA spectral quantities display finite-size scaling properties associated with active-absorbing phase transitions on directed bond percolation and branching-annihilating random walks, analyzing the normalized leading eigenvalue
\begin{equation}
    \pi_1^{u} = \frac{\Lambda_1^{u}}{\sum_i \Lambda_i^{u}},
\end{equation}
where $\Lambda_i^{u}$ denotes the eigenvalues of the uncentered covariance matrix, which encodes the critical behavior. Their finite-size scaling analysis estimated the critical point and critical exponents of the order parameter and correlation length. This demonstrates that the PCA spectrum can encode the scaling properties of nonequilibrium phase transitions beyond dimensional reduction.

We complement these results by constructing PCA observables that are directly related to the order parameter and its fluctuations. The first PCA observable is the ratio of the largest eigenvalues $\Lambda_2/\Lambda_1$, that obeys the finite-size scaling form
\begin{equation}
   \frac{\Lambda_2}{\Lambda_1} = {\cal R}\left[L^{1/\nu_\perp}(\lambda-\lambda_c)\right],
   \label{eq:fss_eigenvalue_ratio}
\end{equation}
as we will show from the simulation results, where $\nu_\perp$ is the spatial correlation-length exponent. From the hypothesized finite-size scaling form in Eq.~\eqref{eq:fss_eigenvalue_ratio}, the ratio $\Lambda_2/\Lambda_1$ should be consistent with a universal value at the critical threshold, and the crossing for different lattice sizes allows for the estimation of the critical threshold.

The second PCA observable is the average absolute value of the first principal component
\begin{equation}
   P_1 = \left\langle |p_1| \right\rangle.
   \label{eq:P1_definition}
\end{equation}
The absolute value is required because the sign-reversed configurations generate equivalent projections with opposite signs. The average $P_1$ displays a direct connection with the order parameter. This correspondence can be understood from the structure of the leading principal direction. If the first PCA eigenvector is associated with the spatially uniform density mode, then, for a square lattice with $N=L^2$, we have
\begin{equation}
   p_1 = \frac{1}{\sqrt{N}}\sum_{i=1}^{N}\eta_i = L \rho,
   \label{eq:pca_density_relation}
\end{equation}
where $\rho$
\begin{equation}
   \rho=\frac{1}{N}\sum_{i=1}^{N}\eta_i,
   \label{eq:rho_definition}
\end{equation}
is the instantaneous density of active sites. Hence, $P_1/L$ provides a PCA-based estimator of the order parameter $\langle\rho\rangle$, as tested in our simulations. Accordingly, the first principal component is expected to follow the same
finite-size scaling behavior as the order parameter,
\begin{equation}
   \frac{P_1}{L} = L^{-\beta/\nu_\perp} {\cal M}\left[L^{1/\nu_\perp}(\lambda-\lambda_c)\right],
   \label{eq:fss_pca_order}
\end{equation}
where $\beta$ is the order parameter exponent.

The third PCA observable is a fluctuation-like quantity that can also be constructed from the largest PCA eigenvalue. Since the variance along the first principal direction is given by $\Lambda_1$, we define
\begin{equation}
   \chi_{\rm PCA} = \Lambda_1-P_1^2.
   \label{eq:pca_fluctuation}
\end{equation}
For the sign-symmetrized dataset, $\Lambda_1=\langle p_1^2\rangle$. Therefore,
\begin{equation}
 \chi_{\rm PCA} = \left\langle p_1^2\right\rangle - \left\langle |p_1|\right\rangle^2,
\end{equation}
and finally, from Eq.~\eqref{eq:pca_density_relation}, this quantity becomes
\begin{equation}
   \chi_{\rm PCA} = N \left(\langle\rho^2\rangle-\langle\rho\rangle^2\right)
   \label{eq:pca_susceptibility_relation}
\end{equation}
which has the same form as the fluctuations of the order parameter. We therefore expect the following scaling behavior 
\begin{equation}
   \Lambda_1-P_1^2 = L^{\gamma'/\nu_\perp} {\cal X}\left[L^{1/\nu_\perp}(\lambda-\lambda_c)\right],
   \label{eq:fss_pca_fluctuation}
\end{equation}
where $\gamma'$ is the fluctuation exponent associated with the order parameter.

\begin{figure}[pos=h!]
   \begin{center}
   \includegraphics[scale=0.33]{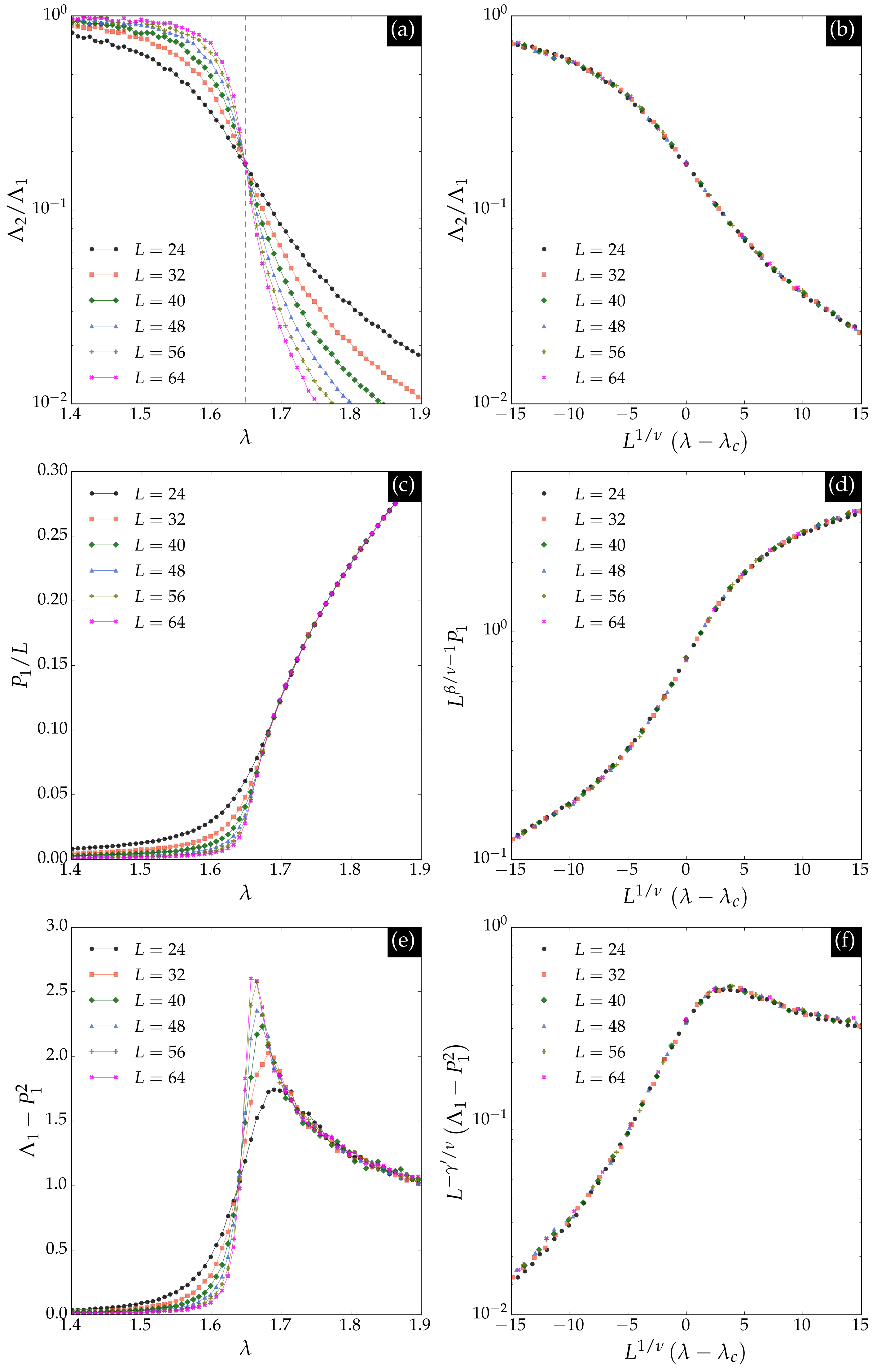}
   \end{center}
   \caption{Principal component analysis of the contact process on the square lattice for different linear system sizes $L$.
   (a) Ratio between the two largest eigenvalues, $\Lambda_2/\Lambda_1$, as a function of $\lambda$. The dashed vertical line indicates $\lambda_c=1.64877$. (b) Finite-size collapse of $\Lambda_2/\Lambda_1$ as a function of $L^{1/\nu_\perp}(\lambda-\lambda_c)$. (c) PCA-based order parameter $P_1/L$. The PCA-based order parameter coincides with the density of active sites. (d) Corresponding finite-size collapse, $L^{\beta/\nu_\perp-1}P_1$. (e) PCA fluctuation $\Lambda_1-P_1^2$. (f) Corresponding finite-size collapse, $L^{-\gamma'/\nu_\perp}(\Lambda_1-P_1^2)$. The numerical values of critical exponents are $\nu_\perp=0.7333$, $\beta=0.5834$, and $\gamma'=0.2998$\cite{henkel-2008}.}
   \label{fig:pca_square}
\end{figure}

Fig.~\ref{fig:pca_square} summarizes the PCA results for the contact process on the square lattice. We separate the configurations by its control parameter $\lambda$ and apply PCA separately by each value of the control parameter, therefore all observables are functions of $\lambda$. Panel~(a) of Fig.~\ref{fig:pca_square} shows the ratio $\Lambda_2/\Lambda_1$ between the two largest PCA eigenvalues. In the low-$\lambda$ regime the two leading eigenvalues are comparable, whereas the ratio decreases rapidly upon entering the active phase, indicating the increasing dominance of the first principal direction. As the system size increases, the crossover becomes progressively sharper in the vicinity of the critical threshold. The ratio is consistent with a universal value at the critical threshold as already discussed, so no vertical rescaling is required. Panel~(b) shows $\Lambda_2/\Lambda_1$ as a function of the scaling variable $L^{1/\nu_\perp}(\lambda-\lambda_c)$ revealing an excellent collapse of the curves for different system sizes.

As shown in Fig.~\ref{fig:pca_square}(c), the quantity $P_1/L$ is the density of active sites: it approaches zero in the absorbing phase and becomes finite in the active phase, characteristic of a continuous phase transition. The collapse shown in panel~(d) demonstrates that the PCA-based order parameter follows the expected critical scaling. Panel~(e) exhibits the corresponding fluctuation peak, whose height increases and whose width decreases with increasing $L$. The collapse shown in panel~(f) confirms that this PCA-derived quantity follows the finite-size scaling expected for the fluctuations of the order parameter.

For the square lattice, the collapses in Fig.~\ref{fig:pca_square} were obtained using $\lambda_c=1.64877$, $\nu_\perp=0.7333$,
$\beta=0.5834$, $\gamma'=0.2998$, which correspond to $1/\nu_\perp\simeq1.36$, $\beta/\nu_\perp\simeq0.80$, and $\gamma'/\nu_\perp\simeq0.41$. The simultaneous collapse of the dimensionless eigenvalue ratio, the PCA-based order parameter, and its fluctuation provides strong evidence that, after the sign-reversed data augmentation, PCA recovers not only the location of the absorbing-state transition but also the finite-size scaling structure of the underlying directed-percolation critical threshold. We report that similar behavior also happens in the contact process in the one-dimensional chain where the critical exponents are drawn from the one-dimensional directed percolation universality class and in the triangular lattice.

\section{Heteroscedastic Variational Autoencoder}\label{sec:vae_results}

We next investigate whether a generative unsupervised deep learning model can identify the active-absorbing phase transition directly from microscopic contact process configurations. To this end, we employ a variational autoencoder (VAE)~\cite{alexandrou-2020,shen-2021,shen-2022}, a generative model consisting of an encoder that maps input configurations onto a low-dimensional latent distribution and a decoder that generates reconstructed configurations from latent samples. The model is a conditional VAE, where both the encoder and decoder are conditioned on the control parameter $\lambda$.

For a microscopic input configuration $\boldsymbol{\eta}_{\mathrm{real}}$, the encoder parametrizes the approximate posterior distribution as
\begin{equation}
   q_{\phi}\left(\boldsymbol{z}\mid\boldsymbol{\eta}_{\mathrm{real}}\right) =
   \mathcal{N}\left[\boldsymbol{\mu},\operatorname{diag}\left(\boldsymbol{\sigma}^{2}\right) \right],
   \label{eq:vae-posterior}
\end{equation}
where $\boldsymbol{\mu}$ and $\boldsymbol{\sigma}^{2}$ are encoder outputs and $\phi$ denotes trainable parameters. A latent configuration $\boldsymbol{z}$ is sampled from this distribution and fed to the decoder, which produces a reconstructed configuration $\boldsymbol{\eta}_{\mathrm{recon}}$.

Standard VAEs regularize the latent distribution using a fixed standard Gaussian prior, $p(\boldsymbol{z})=\mathcal{N}(\boldsymbol{0},\boldsymbol{I})$. However, this prior does not account for the asymmetric and non-negative structure of contact process configurations, particularly critical near the absorbing-state transition where the density of active sites changes continuously from dilute to active regimes.

To address this limitation, we replace the standard Gaussian prior with a parameter-dependent heteroscedastic Gaussian prior,
\begin{equation}
   p_{\theta}\left(\boldsymbol{z}\mid\lambda\right) =
   \mathcal{N} \left[\boldsymbol{\mu}'(\lambda),\operatorname{diag}\left({\boldsymbol{\sigma}'}^{\,2}(\lambda)\right)\right].
   \label{eq:heteroscedastic-prior}
\end{equation}
Both the mean $\boldsymbol{\mu}'(\lambda)$ and variance ${\boldsymbol{\sigma}'}^{2}(\lambda)$ are therefore learned as functions of the control parameter $\lambda$. In contrast to the standard prior, the location and width of the latent distribution are allowed to evolve across the absorbing-state transition.

The encoder receives the $L^2$ binary variables $\eta_i\in\{0,1\}$ defining a contact process configuration. It consists of three fully connected hidden layers containing $625$, $256$, and $64$ neurons, respectively, with ReLU activation functions. Each hidden layer is followed by batch normalization and dropout with probability $p=0.2$. The encoder output parametrizes the mean $\boldsymbol{\mu}$ and logarithm of the variance of the latent distribution. Both the encoder and the decoder are conditioned on the control parameter $\lambda$, which is concatenated to the input of the first layer of each network. The decoder mirrors the encoder architecture and maps latent samples to a reconstructed configuration with sigmoid activations.

The heteroscedastic prior is generated independently by an additional fully connected neural network. This network also receives the control parameter $\lambda$ as input, contains one hidden layer with $32$ neurons, and outputs $\boldsymbol{\mu}'(\lambda)$ and the logarithm of ${\boldsymbol{\sigma}'}^{2}(\lambda)$. All networks were trained using batches of $128$ configurations and the Adam optimizer with learning rate $10^{-3}$ and a ratio 80/20 between training and validation data.

The neural networks were implemented in Python using TensorFlow and its integrated Keras high-level API~\cite{abadi-2016,chollet-2015}. The Keras functional API was used to construct the encoder, decoder, conditional heteroscedastic prior, and custom VAE loss layer, while optimization was performed with the Adam optimizer provided by TensorFlow.

The VAE is trained by minimizing a loss function containing a reconstruction term, and a Kullback--Leibler (KL) regularization term,
\begin{equation}
   \ell_{\mathrm{VAE}} = \ell_{\mathrm{RECON}} + \ell_{\mathrm{KL}}.
   \label{eq:vae-loss}
\end{equation}
The reconstruction term $\ell_{\mathrm{RECON}}$ quantifies how accurately the decoder reproduces the microscopic input configuration. The second contribution to the loss is the KL divergence $\ell_{\mathrm{KL}}$ between the distribution learned by the encoder and the heteroscedastic prior. 

For the reconstruction term, one can consider either the mean-squared error (MSE) or the binary cross entropy (BCE). For a dataset containing $N_{\mathcal D}$ configurations, each with $N=L^2$ sites, the MSE is
\begin{equation}
   \ell_{\mathrm{MSE}} = 
   \frac{1}{N_{\mathcal D}N} \sum_{\alpha=1}^{N_{\mathcal D}}\sum_{i=1}^{N}\left[\eta^{(\alpha)}_{i,\mathrm{real}} - \eta^{(\alpha)}_{i,\mathrm{recon}}\right]^2,
   \label{eq:vae-mse}
\end{equation}
whereas the BCE is
\begin{equation}
   \ell_{\mathrm{BCE}} =
   -\frac{1}{N_{\mathcal D}N} \sum_{\alpha=1}^{N_{\mathcal D}} \sum_{i=1}^{N} 
   \left[ \eta^{(\alpha)}_{i,\mathrm{real}} \ln \eta^{(\alpha)}_{i,\mathrm{recon}}
   +\left(1-\eta^{(\alpha)}_{i,\mathrm{real}}\right) \ln \left(1-\eta^{(\alpha)}_{i,\mathrm{recon}}\right) \right].
   \label{eq:vae-bce}
\end{equation}
In the present work, we used the MSE reconstruction loss rescaled by $N$ (i.e., $\ell_{\mathrm{RECON}} \equiv N \ell_{\mathrm{MSE}}$) as the reconstruction loss.

The KL regularization term for diagonal Gaussian distributions can be evaluated analytically as
\begin{equation}
   \ell_{\mathrm{KL}} = 
   \frac{1}{2} \sum_{i=1}^{d} 
   \left[ \ln \frac{{\sigma_i'}^{\,2}}{\sigma_i^2} - 1 + \frac{\sigma_i^2}{{\sigma_i'}^{\,2}} + \frac{(\mu_i-\mu_i')^2}{{\sigma_i'}^{\,2}} \right],
   \label{eq:vae-kl}
\end{equation}
where $d$ is the dimension of the latent space (we used $d=1$ in our results), $(\mu_i,\sigma_i^2)$ are the parameters inferred from the microscopic configuration, and $(\mu_i'(\lambda),{\sigma_i'}^{2}(\lambda))$ are the corresponding parameters of the heteroscedastic prior.

\begin{figure}[pos=h!]
   \begin{center}
      \includegraphics[scale=0.33]{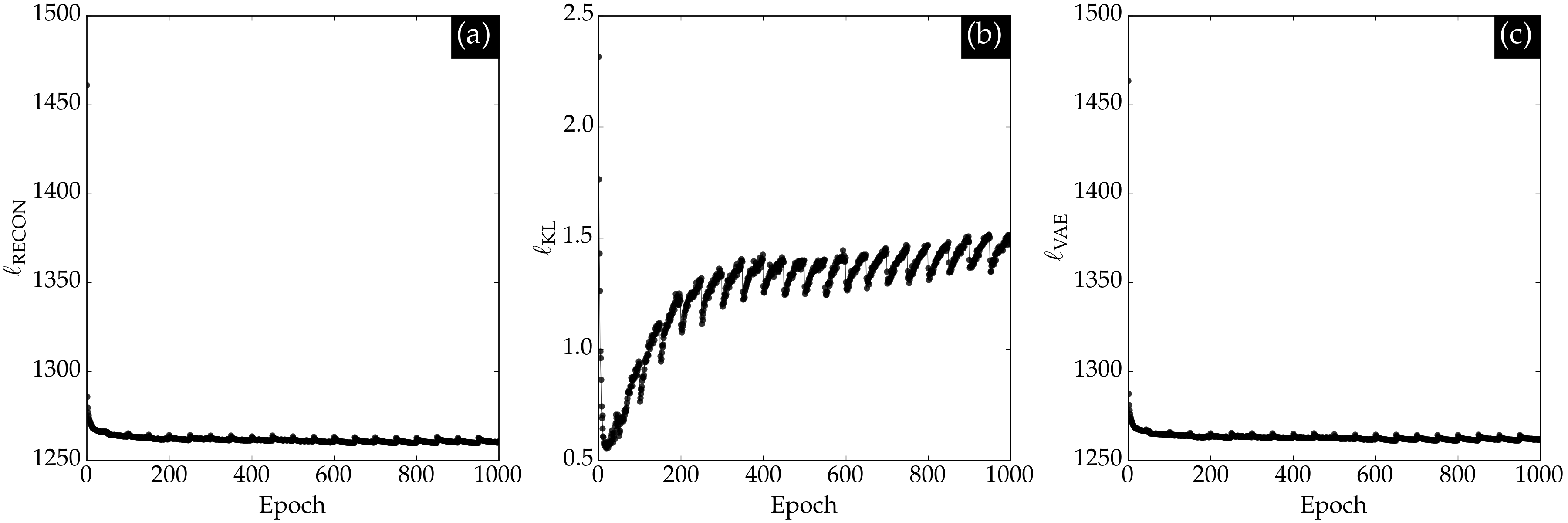}
   \end{center}
   \caption{Training history of the heteroscedastic VAE for the contact process on the square lattice with $L=64$. Panels (a)--(c) show the reconstruction loss $\ell_{\mathrm{RECON}}$, the Kullback--Leibler divergence $\ell_{\mathrm{KL}}$, and the total loss $\ell_{\mathrm{VAE}}=\ell_{\mathrm{RECON}}+\ell_{\mathrm{KL}}$, respectively. The reconstruction loss decreases rapidly before reaching a slowly varying regime, while the Kullback--Leibler contribution remains finite. Small periodic variations arise from training 20 non-overlapping dataset chunks sequentially for 50 epochs each.}
   \label{fig:vae_training}
\end{figure}

Fig.~\ref{fig:vae_training} shows the evolution of the different contributions to the heteroscedastic VAE loss during training on the square lattice with $L=64$. The reconstruction loss decreases rapidly during the initial epochs and then approaches a slowly varying stationary regime, indicating that the reconstruction task has converged. The Kullback--Leibler contribution remains finite throughout training, suggesting that the latent representation retains a nontrivial stochastic character rather than collapsing to a purely deterministic encoding. The small periodic variations in the loss curves result from processing the training set as a sequence of non-overlapping configuration subsets, with each subset used for a fixed number of epochs before being replaced by the next. Despite these fluctuations, the reconstruction and total VAE losses remain stable at long training times, indicating that the optimization reaches a well-defined regime.

To quantify the amount of microscopic information preserved by the generative model, we compute the normalized correlation between the input and reconstructed configurations,
\begin{equation}
   C\left(\boldsymbol{\eta}_{\mathrm{real}} \mid \boldsymbol{\eta}_{\mathrm{recon}} \right)
   \equiv
   \frac{1}{N}
   \frac{\left\langle\left|\boldsymbol{\eta}_{\mathrm{real}}\cdot\boldsymbol{\eta}_{\mathrm{recon}}\right|\right\rangle}
   {\rho_{\mathrm{real}}\rho_{\mathrm{recon}}},
   \label{eq:vae-correlation}
\end{equation}
where $\rho_{\mathrm{real}}$ and $\rho_{\mathrm{recon}}$ are the mean densities of active sites in the original and reconstructed configurations, respectively, computed according to the definition in Eq.~\eqref{eq:rho_definition}. At the critical threshold, the simulation result for $C$ is consistent with a universal dimensionless quantity; consequently, the crossing of the curves for different system sizes can be used to locate the critical point. Its corresponding finite-size scaling form is
\begin{equation}
   C\left(\boldsymbol{\eta}_{\mathrm{real}}\mid\boldsymbol{\eta}_{\mathrm{recon}}\right)-1 =
   \mathcal{C}_{\mathrm{VAE}}\left[L^{1/\nu_\perp}(\lambda-\lambda_c)\right].
   \label{eq:vae-correlation-fss}
\end{equation}

The reconstruction errors $\ell_{\mathrm{MSE}}$ and $\ell_\mathrm{BCE}$ also display critical finite-size behavior. We hypothesize the corresponding \textit{ad hoc} scaling forms as
\begin{equation}
   \begin{split}
      \ell_{\mathrm{MSE}} &= L^{-\beta/\nu_\perp} \mathcal{E}_{\mathrm{MSE}} \left[L^{1/\nu_\perp} (\lambda-\lambda_c) \right], \\
      \ell_{\mathrm{BCE}} &= \left(\frac{L}{\ln L} \right)^{-\beta/\nu_\perp} \mathcal{E}_{\mathrm{BCE}}\left[L^{1/\nu_\perp}(\lambda-\lambda_c)\right],
   \end{split}
   \label{eq:errors-fss}
\end{equation}
The logarithmic factor in $\ell_{\mathrm{BCE}}$ should be regarded here as an empirical finite-size correction suggested by the numerical collapse, rather than as an independently derived critical exponent. Nevertheless, the fact that both reconstruction losses can be collapsed using the directed-percolation exponents demonstrates that the reconstruction quality of the heteroscedastic VAE retains information about the critical fluctuations of the original stochastic system. The collapse is particularly good in the critical scaling regime, while systematic finite-size corrections become more visible away from the transition, especially deeper in the active phase.

\begin{figure}[pos=h!]
   \begin{center}
      \includegraphics[scale=0.33]{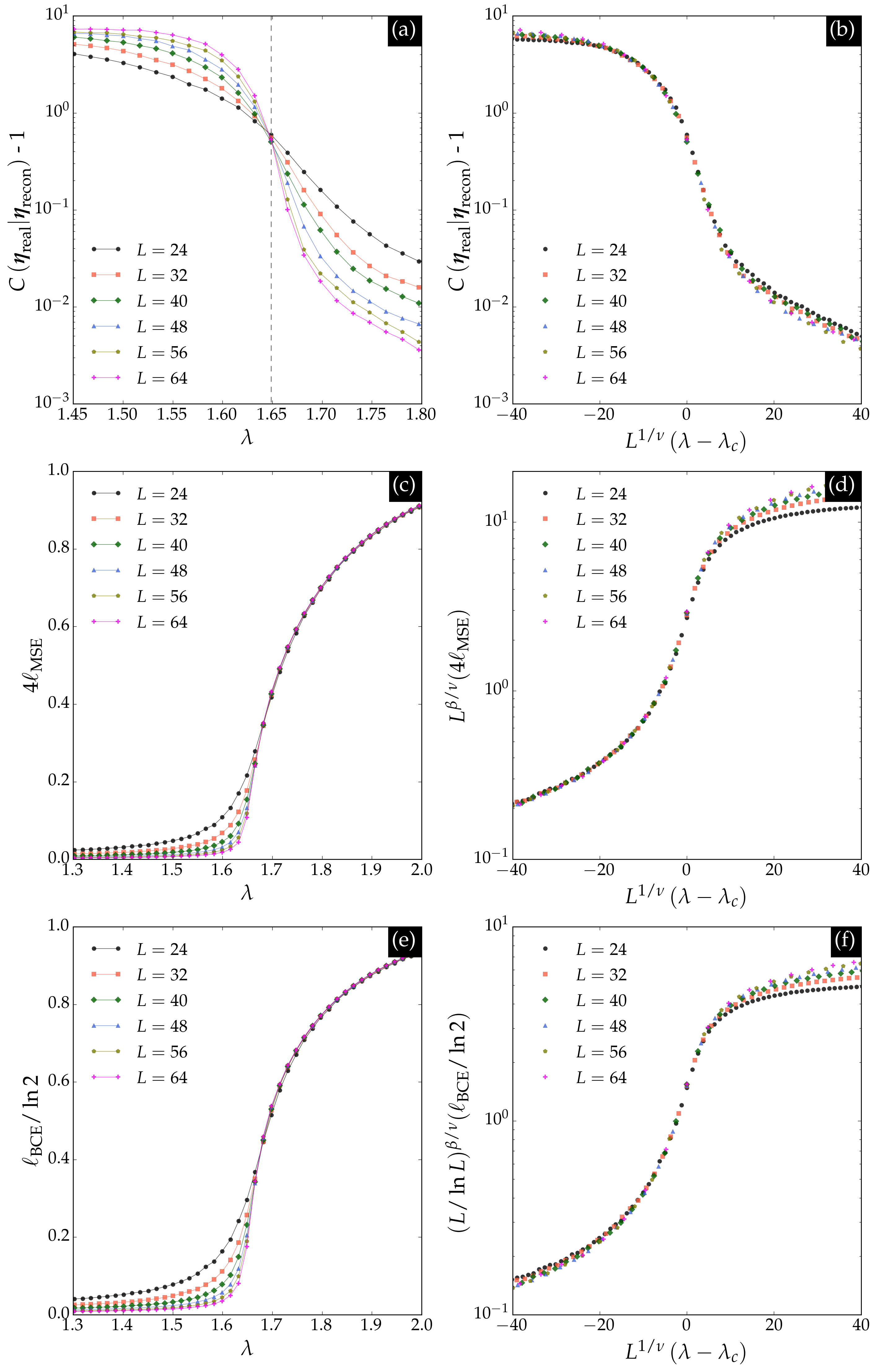}
   \end{center}
   \caption{Heteroscedastic variational autoencoder results for the contact process on the square lattice for different linear system sizes $L$. (a) Excess normalized correlation $C(\boldsymbol{\eta}_{\mathrm{real}}\mid\boldsymbol{\eta}_{\mathrm{recon}})-1$ between the input and reconstructed configurations as a function of $\lambda$. The dashed vertical line marks $\lambda_c=1.64877$. (b) Finite-size collapse of the correlation as a function of $L^{1/\nu_\perp}(\lambda-\lambda_c)$. (c) Normalized mean-squared reconstruction error $\ell_{\mathrm{MSE}}$. (d) Corresponding finite-size scaling, $L^{\beta/\nu_\perp}(4\ell_{\mathrm{MSE}})$. (e) Normalized binary-cross-entropy reconstruction error $\ell_{\mathrm{BCE}}/ \ln 2$. (f) Corresponding collapse including the logarithmic finite-size correction, $(L/\ln L)^{\beta/\nu_\perp}\ell_{\mathrm{BCE}}/\ln 2$. The collapses use $\nu_\perp=0.7333$ and $\beta=0.5834$.}
   \label{fig:vae_square}
\end{figure}

We show in panel (a) of Fig.~\ref{fig:vae_square}, the excess normalized correlation $C(\boldsymbol{\eta}_{\mathrm{real}}\mid\boldsymbol{\eta}_{\mathrm{recon}})-1$ as a function of $\lambda$. The normalized excess correlation decreases rapidly across the transition, and the crossover becomes increasingly sharp as $L$ increases. Although the curves display a strong size dependence when plotted directly as functions of $\lambda$, they collapse onto a common scaling curve. The collapse in panel (b) of Fig.~\ref{fig:vae_square} therefore indicates that the input reconstruction correlation behaves as a universal quantity in the critical region.

In addition, we report that the similar observable $C(\boldsymbol{\eta}_{\mathrm{real}}\mid\boldsymbol{\eta}_{\mathrm{recon}})$ was obtained using a standard VAE. The results are in agreement with those shown here only for the smallest system sizes, failing to cross at the critical threshold for larger lattices. This indicates that the fixed prior is too restrictive to capture the statistical structure of the contact process configurations, and that the heteroscedastic prior is essential to recover the correct scaling behavior.

Panel~(c) of Fig.~\ref{fig:vae_square} displays the rescaled mean-squared reconstruction error $\ell_{\mathrm{MSE}}$. The curves exhibit an increasingly sharp increase near $\lambda_c$, resembling the behavior of an order parameter. The data collapse in panel~(d) of Fig.~\ref{fig:vae_square} is obtained from the scaling form in Eq.~\ref{eq:errors-fss}, and shows that $\ell_{\mathrm{MSE}}$ scaling behavior is consistent with the scaling of a order parameter. A similar behavior emerges for the binary cross-entropy reconstruction error in panel~(e) of Fig.~\ref{fig:vae_square}. However, the optimal collapse of the BCE data shown in panel~(f) of Fig.~\ref{fig:vae_square} requires a logarithmic finite-size correction, as indicated by the empirical expression in Eq.~\ref{eq:errors-fss}. An empirical logarithmic correction substantially improves the collapse, although its theoretical origin remains unclear.

For the square lattice, the finite-size collapses were obtained using the same critical ratios as in the PCA analysis. The same critical ratios provide a consistent collapse for the scaling of the excess normalized correlation and of both reconstruction losses. This agreement indicates that the heteroscedastic VAE encodes the finite-size critical structure of the contact process. In addition, we obtained similar results for the one-dimensional chain where the critical ratios are drawn from the one-dimensional directed percolation universality class. We also report that the same model trained with data on the square lattice was able to recover the critical threshold of the triangular lattice as well when using inference data of the triangular lattice.

\section{Conclusions} \label{sec:conclusions}

In this work, we investigated the ability of unsupervised machine-learning methods to characterize the absorbing-state phase transition of the contact process directly from microscopic configurations. We focused on two complementary approaches, principal component analysis (PCA) and variational autoencoders (VAEs), and found that the direct application of their standard formulation is strongly affected by the non-negative structure of the contact process configuration space.

For PCA, conventional centering of the original configurations does not reconstruct the order parameter. We showed that this limitation can be overcome by augmenting the dataset with the sign-reversed counterpart of each configuration. This artificial symmetrization, introduced exclusively as a preprocessing operation and not as a physical symmetry of the model, produces an exactly centered dataset and allows the leading principal component to recover the behavior of the density of active sites. More importantly, quantities constructed exclusively from the PCA spectrum and the principal components display the finite-size behavior expected at the absorbing-state transition. The corresponding finite-size collapses are consistent with the directed percolation critical behavior.

A complementary picture emerges from the generative deep learning approach. We found that a standard VAE is not sufficiently flexible to represent the statistical structure of the contact process configurations. We therefore introduced a heteroscedastic Gaussian prior whose mean and variance are learned as functions of the control parameter $\lambda$. Consequently, the Kullback--Leibler regularization does not constrain configurations throughout the entire phase diagram to the same reference distribution. Instead, the prior itself adapts to the configurations in the absorbing and active phases. Also, the reconstruction losses exhibit the critical finite-size behavior for the directed percolation universality class.

Taken together, the PCA and VAE results point to a common conclusion: the success of unsupervised learning in systems with absorbing states depends not only on the choice of the learning algorithm, but also on how the statistical structure of the microscopic configuration space is incorporated into the method. In PCA, this is achieved by modifying the representation of the input data through sign-reversed augmentation, whereas in the VAE it is achieved through a parameter-dependent heteroscedastic prior. In both cases, once this structure is properly taken into account, quantities generated by the machine-learning models reproduce the characteristic finite-size scaling of the underlying nonequilibrium phase transition.

These results suggest that adapting unsupervised-learning models to the symmetries, constraints, and statistical support of the microscopic data may provide a general route for studying nonequilibrium systems whose configuration spaces differ substantially from the symmetry-balanced data commonly encountered in equilibrium spin models. Extensions to other absorbing-state models, lattice geometries, and universality classes constitute natural directions for future investigation.

\section*{Funding}
We thank CNPq, CAPES, FAPEPI, and FINEP for their financial support. R. S. Ferreira acknowledges support from FAPEMIG (grant FAPEMIG-APQ-06611-24).

\section*{Author Contribution Statement}
All authors contributed equally to the research conceptualization, result analysis, and manuscript writing.

\section*{Data Availability Statement}
The datasets generated and analyzed during the current study are available from the corresponding author upon reasonable request.

\bibliographystyle{unsrtnat}
\bibliography{textv1}

@article{harris-1974,
  author  = {Harris, T. E.},
  title   = {Contact Interactions on a Lattice},
  journal = {The Annals of Probability},
  volume  = {2},
  number  = {6},
  pages   = {969--988},
  year    = {1974},
  doi     = {10.1214/aop/1176996493}
}

@article{hinrichsen-2000,
  author  = {Hinrichsen, Haye},
  title   = {Non-equilibrium critical phenomena and phase transitions into absorbing states},
  journal = {Advances in Physics},
  volume  = {49},
  number  = {7},
  pages   = {815--958},
  year    = {2000},
  doi     = {10.1080/00018730050198152}
}

@article{odor-2004,
  author  = {\'{O}dor, G\'{e}za},
  title   = {Universality classes in nonequilibrium lattice systems},
  journal = {Reviews of Modern Physics},
  volume  = {76},
  number  = {3},
  pages   = {663--724},
  year    = {2004},
  doi     = {10.1103/RevModPhys.76.663}
}

@book{henkel-2008,
  author    = {Henkel, Malte and Hinrichsen, Haye and L\"ubeck, Sven},
  title     = {Non-Equilibrium Phase Transitions: Volume 1: Absorbing Phase Transitions},
  publisher = {Springer},
  address   = {Dordrecht},
  year      = {2008},
  series    = {Theoretical and Mathematical Physics},
  doi       = {10.1007/978-1-4020-8765-3},
  isbn      = {978-1-4020-8765-3}
}

@article{jensen-1992,
  author  = {Jensen, Iwan},
  title   = {Critical behavior of the three-dimensional contact process},
  journal = {Physical Review A},
  volume  = {45},
  number  = {2},
  pages   = {R563--R566},
  year    = {1992},
  doi     = {10.1103/PhysRevA.45.R563}
}

@article{mehta-2019,
  author  = {Mehta, Pankaj and Bukov, Marin and Wang, Ching-Hao and Day, Alexandre G. R. and Richardson, Clint and Fisher, Charles K. and Schwab, David J.},
  title   = {A high-bias, low-variance introduction to Machine Learning for physicists},
  journal = {Physics Reports},
  volume  = {810},
  pages   = {1--124},
  year    = {2019},
  doi     = {10.1016/j.physrep.2019.03.001}
}

@article{carrasquilla-2017,
  author  = {Carrasquilla, Juan and Melko, Roger G.},
  title   = {Machine learning phases of matter},
  journal = {Nature Physics},
  volume  = {13},
  pages   = {431--434},
  year    = {2017},
  doi     = {10.1038/nphys4035}
}

@article{wang-2016,
  author  = {Wang, Lei},
  title   = {Discovering phase transitions with unsupervised learning},
  journal = {Physical Review B},
  volume  = {94},
  number  = {19},
  pages   = {195105},
  year    = {2016},
  doi     = {10.1103/PhysRevB.94.195105}
}

@article{wetzel-2017,
  author  = {Wetzel, Sebastian J.},
  title   = {Unsupervised learning of phase transitions: From principal component analysis to variational autoencoders},
  journal = {Physical Review E},
  volume  = {96},
  number  = {2},
  pages   = {022140},
  year    = {2017},
  doi     = {10.1103/PhysRevE.96.022140}
}

@article{hu-2017,
  author  = {Hu, Wenjian and Singh, Rajiv R. P. and Scalettar, Richard T.},
  title   = {Discovering phases, phase transitions, and crossovers through unsupervised machine learning: A critical examination},
  journal = {Physical Review E},
  volume  = {95},
  number  = {6},
  pages   = {062122},
  year    = {2017},
  doi     = {10.1103/PhysRevE.95.062122}
}

@article{shen-2021,
  author  = {Shen, Jianmin and Li, Wei and Deng, Shengfeng and Zhang, Tao},
  title   = {Supervised and unsupervised learning of directed percolation},
  journal = {Physical Review E},
  volume  = {103},
  number  = {5},
  pages   = {052140},
  year    = {2021},
  doi     = {10.1103/PhysRevE.103.052140}
}

@article{shen-2022,
  author  = {Shen, Jianmin and Li, Wei and Deng, Shengfeng and Xu, Dian and Chen, Shiyang and Liu, Feiyi},
  title   = {Machine learning of pair-contact process with diffusion},
  journal = {Scientific Reports},
  volume  = {12},
  pages   = {19728},
  year    = {2022},
  doi     = {10.1038/s41598-022-23350-2}
}

@article{tirelli-2022,
  author  = {Tirelli, Andrea and Carvalho, Danyella O. and Oliveira, Lucas A. and de Lima, Jos\'{e} P. and Costa, Natanael C. and dos Santos, Raimundo R.},
  title   = {Unsupervised machine learning approaches to the q-state Potts model},
  journal = {The European Physical Journal B},
  volume  = {95},
  pages   = {189},
  year    = {2022},
  doi     = {10.1140/epjb/s10051-022-00453-3}
}

@article{yevick-2022,
  author  = {Yevick, David},
  title   = {Variational autoencoder analysis of Ising model statistical distributions and phase transitions},
  journal = {The European Physical Journal B},
  volume  = {95},
  pages   = {56},
  year    = {2022},
  doi     = {10.1140/epjb/s10051-022-00296-y}
}

@article{gao-2025,
  author  = {Gao, Feng and Shen, Jianmin and Wang, Shanshan and Li, Wei and Xu, Dian},
  title   = {Neural network learning of multi-scale and discrete temporal features in directed percolation},
  journal = {Physics Letters A},
  volume  = {563},
  pages   = {131071},
  year    = {2025},
  doi     = {10.1016/j.physleta.2025.131071}
}

@article{pruessner-2007,
  author  = {Pruessner, Gunnar},
  title   = {Equivalence of conditional and external field ensembles in
             absorbing-state phase transitions},
  journal = {Physical Review E},
  volume  = {76},
  number  = {6},
  pages   = {061103},
  year    = {2007},
  doi     = {10.1103/PhysRevE.76.061103}
}

@article{macedo-filho-2018,
  author  = {Macedo-Filho, A. and Alves, G. A. and
             Costa Filho, R. N. and Alves, T. F. A.},
  title   = {Reactivating dynamics for the susceptible-infected-susceptible
             model: A simple method to simulate the absorbing phase},
  journal = {Journal of Statistical Mechanics: Theory and Experiment},
  volume  = {2018},
  number  = {4},
  pages   = {043208},
  year    = {2018},
  doi     = {10.1088/1742-5468/aab04a}
}

@article{pedregosa-2011,
  author  = {Pedregosa, Fabian and Varoquaux, Ga{\"e}l and Gramfort, Alexandre
             and Michel, Vincent and Thirion, Bertrand and Grisel, Olivier
             and Blondel, Mathieu and Prettenhofer, Peter and Weiss, Ron
             and Dubourg, Vincent and Vanderplas, Jake and Passos, Alexandre
             and Cournapeau, David and Brucher, Matthieu and Perrot, Matthieu
             and Duchesnay, {\'E}douard},
  title   = {Scikit-learn: Machine Learning in Python},
  journal = {Journal of Machine Learning Research},
  volume  = {12},
  pages   = {2825--2830},
  year    = {2011}
}

@article{alexandrou-2020,
  author  = {Alexandrou, Constantia and Athenodorou, Andreas
             and Chrysostomou, Charalambos and Paul, Srijit},
  title   = {The critical temperature of the 2D-Ising model through
             deep learning autoencoders},
  journal = {The European Physical Journal B},
  volume  = {93},
  number  = {12},
  pages   = {226},
  year    = {2020},
  doi     = {10.1140/epjb/e2020-100506-5}
}

@inproceedings{abadi-2016,
  author    = {Abadi, Mart{\'i}n and Barham, Paul and Chen, Jianmin
               and Chen, Zhifeng and Davis, Andy and Dean, Jeffrey
               and Devin, Matthieu and Ghemawat, Sanjay and Irving, Geoffrey
               and Isard, Michael and Kudlur, Manjunath and Levenberg, Josh
               and Monga, Rajat and Moore, Sherry and Murray, Derek G.
               and Steiner, Benoit and Tucker, Paul and Vasudevan, Vijay
               and Warden, Pete and Wicke, Martin and Yu, Yuan
               and Zheng, Xiaoqiang},
  title     = {TensorFlow: A System for Large-Scale Machine Learning},
  booktitle = {12th USENIX Symposium on Operating Systems Design
               and Implementation (OSDI 16)},
  pages     = {265--283},
  year      = {2016},
  publisher = {USENIX Association}
}

@misc{chollet-2015,
  author = {Chollet, Fran{\c{c}}ois and others},
  title  = {Keras},
  year   = {2015},
  note   = {Deep learning library for Python}
}

@article{neto-2025,
  author  = {Neto, J. F. S. and Alencar, D. S. M. and Brito, L. T.
             and Alves, G. A. and Lima, F. W. S. and Macedo-Filho, A.
             and Ferreira, R. S. and Alves, T. F. A.},
  title   = {Deep Learning of the Biswas--Chatterjee--Sen Model},
  journal = {Entropy},
  volume  = {27},
  number  = {11},
  pages   = {1173},
  year    = {2025},
  doi     = {10.3390/e27111173}
}

@article{neto-2026,
  author  = {Neto, J. F. S. and Alencar, D. S. M. and Brito, L. T.
             and Alves, G. A. and Lima, F. W. S. and Macedo-Filho, A.
             and Ferreira, R. S. and Alves, T. F. A.},
  title   = {Supervised and Unsupervised Deep Learning Applied to the
             Majority Vote Model},
  journal = {Physica A: Statistical Mechanics and its Applications},
  volume  = {683},
  pages   = {131208},
  year    = {2026},
  doi     = {10.1016/j.physa.2025.131208}
}

@article{muzzi-2024,
  author  = {Muzzi, Cristiano and Cortes, Ronald Santiago
             and Bhakuni, Devendra Singh and Jeli{\'c}, Asja
             and Gambassi, Andrea and Dalmonte, Marcello
             and Verdel, Roberto},
  title   = {Principal component analysis of absorbing state phase transitions},
  journal = {Physical Review E},
  volume  = {110},
  number  = {6},
  pages   = {064121},
  year    = {2024},
  doi     = {10.1103/PhysRevE.110.064121}
}

\end{document}